\documentclass[twocolumn,aps,prd,preprintnumbers]{revtex4}

\usepackage{graphicx}
\usepackage{dcolumn}
\usepackage{bm}
\usepackage{subfigure}
\usepackage{slashed}
\usepackage{marvosym}
\usepackage{textcomp}
\usepackage{threeparttable}
\usepackage{booktabs}
\usepackage{setspace}
\usepackage{latexsym}
\usepackage{mciteplus}
\renewcommand{\TPTtagStyle}%
{\normalsize\textit}

\begin{document}

{\raggedleft{}CCTP-2015-16, CCQCN-2015-97}
\title{Shear Viscosities of Photons in Strongly Coupled Plasmas

}
\author{Di-Lun Yang$^{1,2}$\footnote{dilunyang@gmail.com}, Berndt M\"uller$^{3,4}$\footnote{muller@phy.duke.edu}}
\affiliation{
$^1$Theoretical Research Division, Nishina Center, RIKEN, Wako, Saitama 351-0198, Japan.\\
$^2$Crete Center for Theoretical Physics, Department of Physics
University of Crete, 71003 Heraklion, Greece.\\
$^3$Department of Physics, Duke University, Durham, North Carolina 27708, USA.\\
$^4$Brookhaven National Laboratory, Upton, NY 11973, USA.} %
\date{\today}
\begin{abstract}
We investigate the shear viscosity of thermalized photons in the quark gluon plasma (QGP) at weak coupling and $\mathcal{N}=4$ super Yang-Mills  plasma (SYMP) at both strong and weak couplings. We find that the shear viscosity due to the photon-parton scattering up to the leading order of electromagnetic coupling is suppressed when the coupling of the QGP/SYMP is increased, which stems from the blue-shift of the thermal-photon spectrum at strong coupling. In addition, the shear viscosity rapidly increases near the deconfinement transition in a phenomenological model analogous to the QGP.   
\end{abstract}

\maketitle
The electromagnetic (EM) signatures such as leptons and photons produced in thermal plasmas play an important role in probing the properties of the plasmas, which have been widely investigated in relativistic heavy ion collisions and cosmology. 
Peculiarly, recent observations of large elliptic flow of direct photons comparable to that of hadrons in heavy ion collisions \cite{Adare:2011zr,Lohner:2012ct} lead to the tension between experimental results and theoretic predictions \cite{Dion:2011pp,Shen:2013cca}.  
This inconsistency also stimulates many studies for the electromagnetic probes in the QGP with distinct mechanisms \cite{Bzdak:2012fr,Fukushima:2012fg,Muller:2013ila,Linnyk:2013wma,McLerran:2014hza,Gale:2014dfa,Monnai:2014kqa,vanHees:2014ida,Linnyk:2015tha,McLerran:2015mda}. 
Although the photons generated in heavy ion collisions hardly reach thermal equilibrium due to short lifetime of the QGP, some of the photons or leptons produced in the cosmic plasma with longer lifetime may reach equilibrium. In such a case, e.g. in the quark epoch, the cosmic plasma could be regarded as a QED+QCD plasma.

It is found that the quantum electrodynamic (QED) plasma is rather viscous and the thermal leptons or photons barely interact with each other after being emitted from the plasma \cite{Heiselberg:1994vy,Arnold:2000dr,Arnold:2003zc}, whereas the scenario in the QGP is not so obvious since the interaction between the leptons/photons and the medium could be enhanced by the lepton-quark and photon-parton scatterings. Particularly, the photon-parton scattering dominates the photon-photon/lepton interaction in the order of EM coupling. Moreover, in the strongly coupled QGP (sQGP), the non-perturbative effect in the color sector may modify the interaction with the EM sector.   
We thus consider the scenario when leptons/photons and the QGP reach thermal equilibrium with same temperature, while they may not share same transport coefficients due to different couplings in the EM and color sectors. Theoretically, it has been shown recently that the shear viscosity of thermalized leptons in the QGP could be suppressed by the lepton-quark scattering compared with the case in the QED plasma \cite{Muller:2015maa}. Therefore, it is tentative and imperative to analyze the photon transport in the QGP. In this paper, we follow the approach in \cite{Muller:2015maa} to compute the shear viscosity of photons stemming from the photon-parton scattering in the QGP. We apply the relativistic Boltzmann equation to describe the dynamics of thermalized photon, while the collisional terms could be obtained from both perturbative or non-perturbative approaches depending on the coupling of the color sector. However, due to the limitation of lattice simulations for real-time observables, we may resort to the gauge/gravity duality \cite{Maldacena:1997re,Witten:1998qj,Aharony:1999ti}, the correspondence between $d$-dimensional strongly coupled gauge theories in large $N_c$ and $d+1$-dimensional supergravity, to qualitatively delineate the non-perturbative properties of the sQGP. One will see that the corresponding spectral function beyond the hydrodynamic regime at strong coupling yields nontrivial suppression of the shear viscosity of photons.  

Considering only particle scattering, the relativistic Boltzmann equation of photons can be written as 
\begin{eqnarray}\label{Boltzmann_0}
\frac{p^{\mu}}{p^0}\partial_{\mu}f(p,x)=-f(p,x)\Gamma^>(p)+(1+f(p,x))\Gamma^<(p),
\end{eqnarray}  
where 
\begin{eqnarray}
\Gamma^{<(>)}(p)=(2\pi)^3\frac{d\tilde{\Gamma}^{<(>)}}{d^3p}.
\end{eqnarray}
Here $\tilde{\Gamma}^{<(>)}$ represent the production (absorption) rates per spacetime volume and $f(p,x)$ denotes the distribution function of photons in phase space. 
Based on a naive power counting, the electron-quark/electron-scattering cross section is $\mathcal{O}(e^4)$ \footnote{The collinear or infrared divergences actually give rise to the $\mathcal{O}(e^4\ln(1/e))$ contributions.}, whereas the photon-quark/gluon-scattering cross section is $\mathcal{O}(e^2)$. Intuitively, one could expect more substantial influence from the sQGP on photon transport. 
To study the photon transport, we apply the linear response theory to introduce the local fluctuations of photons slightly away from equilibrium in the vicinity of the rest frame but keep the sQGP in equilibrium. Our approach here is distinct from studies of the transport properties of a single system with multiple components such as weakly coupled QGP \cite{Arnold:2000dr,Arnold:2003zc}, where the different quarks and gluons share the same form of fluctuations. In general, the fluctuations of photons also induce the back-reaction on the sQGP, while the back-reaction should be higher-order corrections in $e$. Moreover, by implementing the non-equilibrium fluctuation-dissipation theorem in holography \cite{Mukhopadhyay:2012hv}, one finds that the fluctuations of the production and absorption rates in (\ref{Boltzmann_0}) cancel each other \footnote{Although the non-equilibrium fluctuation-dissipation theorem therein is derived from real scalar fields, it is expected to be applicable for the gauge fields as well.}. Based on the reasons above, we discard the fluctuations of the color sector. In other words, we treat photons as probes; the photon transport here only encodes the equilibrium properties of the sQGP. By introducing $f(p,x)=n_b(p,x)+\delta f(p,x)$ with $n_b(p,x)$ being the thermal distribution of bosons, we find 
\begin{eqnarray}\label{Boltzmann_relax}
\frac{p^{\mu}}{p^0}\partial_{\mu}f(p,x)=\frac{\text{Im}\left[\Pi^R(p^0)\right]}{p^0}\delta f(p,x),
\end{eqnarray}    
where $\Pi^R(p^0)$ denotes the trace of the light-like retarded EM-current correlator of the sQGP.  
To obtain the second equality above, we utilize the following relations,
\begin{eqnarray}
\frac{d\tilde{\Gamma}^<}{d^3p}=\frac{n_b(p,x)}{1+n_b(p,x)}\frac{d\tilde{\Gamma}^>}{d^3p}=-\frac{n_b(p,x)}{(2\pi)^3p^0}\text{Im}\left[\Pi^R(p^0)\right],
\end{eqnarray}
where the photon-production (absorption) rates are associated with the retarded EM-current correlators.
The Boltzmann equation (\ref{Boltzmann_relax}) implies 
\begin{eqnarray}
\tau_{\gamma}=-\left(\frac{\text{Im}\left[\Pi^R(p^0)\right]}{p^0}\right)^{-1}=2p^0\chi(p^0)^{-1},
\end{eqnarray}
where $\tau_{\gamma}$ denotes the relaxation time of photons, which is shown to be inversely proportional to the light-like spectral function $\chi(p^0)$. Note that the relaxation time here depends on the energy of photons. For the low-energy photons, one simply finds $\tau_{\gamma}\approx 1/(2\sigma_c)$, where $\sigma_c$ corresponds to the DC conductivity of the sQGP. 

Now, by taking the shear fluctuation as
\begin{eqnarray}
\delta f(p,x)=(1+n_b(p,x))n_b(p,x)\frac{B(p)}{T}\hat{p}^i\hat{p}^j\partial_{(i}u_{j)},
\end{eqnarray}      
where $\partial_{(i}u_{j)}=\left(\partial_{i}u_{j}+\partial_{j}u_{i}\right)/2$ and $T$ denotes the equilibrium temperature and $u^j$ represents the collective velocity of photons. After solving (\ref{Boltzmann_relax}), we obtain
\begin{eqnarray}
B(p)=-\frac{|{\bf p}|^2}{\text{Im}\left[\Pi^R(p^0)\right]},
\end{eqnarray}
which gives rise to the shear viscosity of photons,
\begin{eqnarray}\label{eta_gamma}\nonumber
\eta_{\gamma}=-\frac{1}{60T\pi^2}\int d|{\bf p}|\frac{|{\bf p}|^5n_b(p,x)(1+n_b(p,x))}{\text{Im}\left[\Pi^R(p^0)\right]}.
\end{eqnarray}
Alternatively, one may rewrite it in terms of the production rate,
\begin{eqnarray}\label{eta_gamma2}
\eta_{\gamma}=\frac{1}{120T\pi^4}\int d|{\bf p}||{\bf p}|^6n_b(p,x)^2(1+n_b(p,x))\left(\frac{d\tilde{\Gamma}^<}{dp}\right)^{-1}.
\end{eqnarray}
In fact, following \cite{Denicol:2011fa}, one can further derive a more general expression of the shear corrections, which takes the form
\begin{eqnarray}
\delta\bar{T}^{ij}(k)=\int\frac{d^3p}{(2\pi)^3p^0}p^ip^j\delta\bar{f}(k)=2\bar{G}^R(k)\partial^{(i}\bar{u}^{j)},
\end{eqnarray}
where the bar corresponds to the Fourier transform with respect to $x$ and $k$ denotes the dual momentum. Here $\bar{G}^R$ is related to the retarded correlation function of the energy stress tensor, which reads
\begin{eqnarray}\label{GR}
\bar{G}^R(k)=\frac{-1}{15}\int\frac{d^3{\bm p}}{(2\pi)^3}\frac{|{\bm p}|^2}{T}\frac{\bar{n}_p(1+\bar{n}_p)}{\left(ik^0-{\bm k}\cdot {\bm v}+\frac{\text{Im}\left(\Pi^{R}(p^0)\right)}{p^0}\right)},
\end{eqnarray}  
where ${\bm v}={\bm p}/|{\bm p}|$. The shear viscosity is defined as $\eta_{\gamma}=\bar{G}^R(0)$, which agrees with the expression in (\ref{eta_gamma2}). According to \cite{Denicol:2010xn,Denicol:2011fa}, one can further study the shear relaxation time $\tau^{\pi}_{\gamma}=-i/(k^0)_1$ with $(k^0)_1$ being the first pole of $\bar{G}^R(k^0,{\bm k}=0)$. Although the shear relaxation time is important for understanding shear corrections, numerically finding the poles from $(\ref{GR})$ is more involved. We thus only focus on the shear viscosity in this paper. 

The general expression in (\ref{eta_gamma2}) is applicable for both the strongly/weakly coupled QGP or different types of media coupled to photons. We focus on the universal feature of $\eta_{\gamma}$ affected by the coupling of the medium. Thus, we will firstly investigate the $\eta_{\gamma}$ in the $\mathcal{N}=4$ super Yang-Mills (SYM) theory despite its difference from QCD, where the EM-current correlators of the SYM plasma in both weakly coupled and strongly couped scenarios have been studied \cite{CaronHuot:2006te}. In addition, the photon emission at finite but large t'Hooft coupling $\lambda$ in the SYM plasma was analyzed in holography \cite{Hassanain:2011ce,Hassanain:2012uj}, which allows us to further explore the change of $\eta_{\gamma}$ in large $\lambda$. The EM-current correlator in the weakly coupled QCD at finite temperature could be found in \cite{Arnold:2000dr,Arnold:2003zc}. Nonetheless, to qualitatively capture the non-conformal effect near the critical temperature in the strongly coupled scenario, we will employ a phenomenological model in holography to mimic the sQGP \cite{Gubser:2008ny,Finazzo:2013efa}. 

We firstly consider the photon transport in the $\mathcal{N}=4$ SYM plasma at strong coupling and large $N_c$ limit, where the thermal-photon production rate was computed in \cite{CaronHuot:2006te}, where two of the Weyl fermions have electric charge $\pm 1/2$ and two complex scalars have electric charge $1/2$ \footnote{The more rigorous computations from the top-down approach could be found in the D3/D7 system \cite{Mateos:2007yp}, where the photons are emitted from quarks in fundamental representation via the embedding of flavor branes. While the physical interpretation therein is distinct from the one in bottom-up approach in \cite{CaronHuot:2006te}, two results only differ by an overall factor in the linear response regime. Further normalization has to be considered wehn comparing with the  electric conductivity in lattice simulations.}. Although the charge assignment is not unique, the corresponding electric conductivity roughly matches the lattice simulation around $T\sim 3.5T_c$ \cite{Amato:2013naa} (see the comparison in \cite{Greif:2014oia}).  
From \cite{CaronHuot:2006te}, the trace of the light-like spectral function reads
\begin{eqnarray}\label{chi_inf}
\frac{\hat{\chi}(p^0)}{\omega}&=&-2\omega^{-1}\text{Im}\left[\hat{\Pi}^R(p^0)\right]
\\\nonumber
&=&\frac{1}{
\Big| {}_2F_1\left(\frac{2-(1+i)\omega}{2},\frac{2+(1-i)\omega}{2};1-i\omega;-1\right)\Big|^{2}},
\end{eqnarray}    
where $\omega=p^0/(2\pi T)$ and the hat corresponds to the normalization by $e^2N_c^2T^2/8$. By using (\ref{eta_gamma}) and (\ref{chi_inf}) and perform the numerical integration of $p$, one finds
\begin{eqnarray}
\eta^{SYM}_{\gamma}(\lambda=\infty)=\frac{1.46T^3}{e^2N_c^2}.
\end{eqnarray}
In comparison with the shear viscosity of electrons $\eta_e$ in the strongly coupled SYM plasmas \cite{Muller:2015maa}, the $\eta_{\gamma}$ is $\mathcal{O}(e^2\ln(1/e))$ suppressed
\footnote{To work in the framework of the D3/D7 system \cite{Mateos:2007yp} as in \cite{Muller:2015maa}, one simply have to replace $e^2N_c^2$ with $4Q^2N_cN_f$.}.
Since we now treat thermal photons as an independent system separated from the SYM plasma,  
it could be intriguing to compare the $\eta^{SYM}_{\gamma}/s_{\gamma}$, where $s_{\gamma}$ denotes the entropy density of photons, with the ratio to shear viscosity and entropy density of the strongly coupled SYM plasma itself, which is equal to $1/(4\pi)$ and known as the Kovtun-Son-
Starinets (KSS) bound \cite{Policastro:2001yc,Kovtun:2004de}. By simply using the entropy density of an ideal photon gas, $s_{\gamma}=4\pi^2T^3/45$, we find
\begin{eqnarray}\label{photon_etas}
\frac{\eta^{SYM}_{\gamma}(\lambda=\infty)}{s_{\gamma}}\approx \frac{25}{4\pi},
\end{eqnarray}
where we take $N_c=3$ and $e^2=\frac{4\pi}{137}$.
Although the ratio is much larger than the one for the SYM plasma itself, it is highly suppressed compared with that in the QED plasma or in the weakly interacting case as we will see. 

\begin{figure}[t]
{\includegraphics[width=7cm,height=4.5cm,clip]{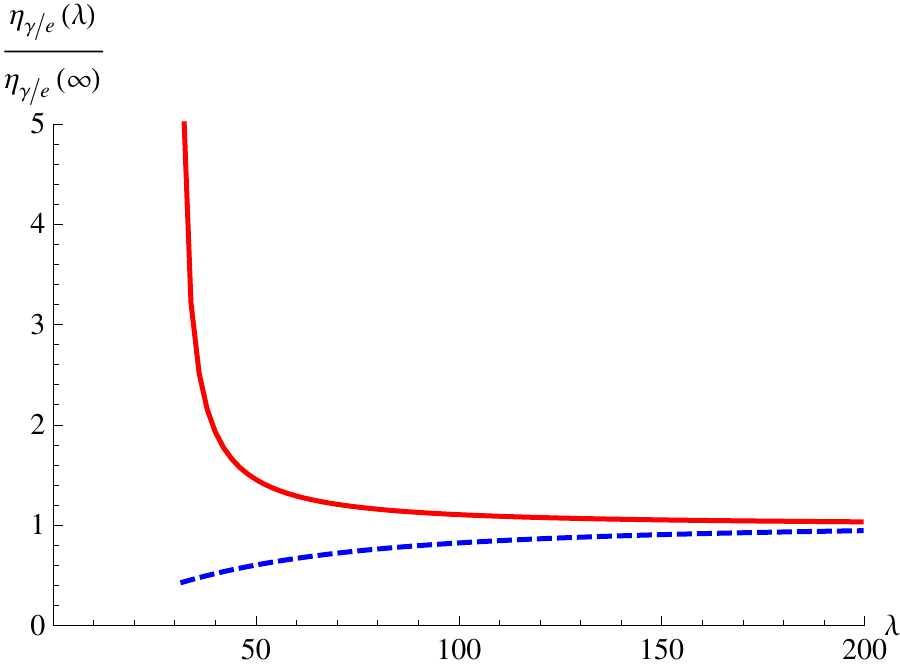}}
\caption{The shear viscosity ratios of photons(red-solid) and leptons(blue-dashed) from holography in the strongly interacting SYM plasma with large t'Hooft coupling. }\label{photon_finite_coupling}
\end{figure}

\begin{figure}[t]
{\includegraphics[width=7cm,height=4.5cm,clip]{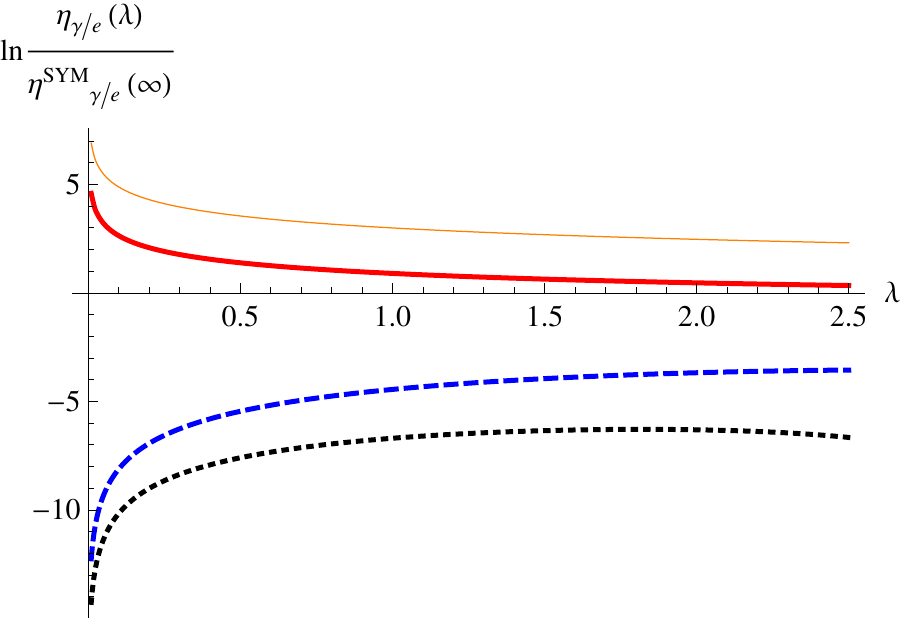}}
\caption{The shear viscosity ratios of photons (red-thick) and leptons (blue-dashed) for the weakly interacting SYM plasma. The shear viscosity of photons (orange-thin) and leptons (black-dotted) in the weakly coupled QGP are also scaled by those in the SYM plasma at $\lambda=\infty$. }\label{photon_weak_coupling}
\end{figure}

Subsequently, we will evaluate $\eta^{SYM}_{\gamma}(\lambda)$ with large but finite $\lambda$, which allows us to track the coupling dependence of $\eta_{\gamma}^{SYM}$. Considering the inclusion of full $\mathcal{O}(\alpha'^3)$ type IIB string theory corrections, the strongly coupled SYM theory receives the $\mathcal{O}(\lambda^{-3/2})$ correction in holography. The thermal-photon production in the SYM plasma with such a correction was studied in \cite{Hassanain:2011ce}. Technically, the computation of the EM-current correlator requires solving a Schr\"odinger equation,
\begin{eqnarray}\label{squ}
\partial_u^2\Psi(u)&=&V(u)\Psi(u),
\\\nonumber
V(u)&=&-\frac{1}{f(u)}
\Bigg(1+\omega^2u-\frac{\gamma}{144}f(u)\Big[-11700
\\\nonumber
&&+2098482u^2-4752055u^4+1838319u^6
\\\nonumber
&&+\omega^2u(-16470+245442u^2+1011173u^4)\Big]\Bigg),
\end{eqnarray}
where $u$ corresponds to the bulk direction with $u=0$ being the boundary and $f(u)=1-\frac{u^2}{u_h^2}$ with $u_h$ being the position of the horizon in the dual geometry. Here $\gamma=\frac{1}{8}\zeta(3)\lambda^{-3/2}$ denotes an expansion parameter with $\zeta$ being the Riemann Zeta function. Eq. (\ref{squ}) has to be solved perturbatively in $\gamma$ with the in-falling boundary condition at the horizon; we thus write 
\begin{eqnarray}
\Psi(u)=(1-u)^{\frac{1}{2}(1-i\omega)}(\Phi_0(u)+\gamma \Phi_1(u)).
\end{eqnarray}
Up to $\mathcal{O}(1)$, solving (\ref{squ}) yields
\begin{eqnarray}
\Phi_0(u)&=&(1+u)^{\frac{1}{2}(1-\omega)}
\\\nonumber
&\times& {}_2F_1\left(\frac{2-(1+i)\omega}{2},\frac{2+(1-i)\omega}{2};1-i\omega;\frac{1-u}{2}\right),
\end{eqnarray}
which is simply the solution giving rise to  (\ref{chi_inf}) at infinite coupling.      
The next-leading-order ($\mathcal{O}(\gamma)$) equation has to be solved numerically. Now the trace of the spectral function is given by
\begin{eqnarray}\label{chi_finite}
\hat{\chi}(p^0)=
4\text{Im}\left[\frac{\Psi_0'}{\Psi_0}+
\gamma\left(\frac{\Psi_1'}{\Psi_0}-\frac{\Psi_0'}{\Psi_0}
\left(\frac{265}{8}+\frac{\Psi_1'}{\Psi_0}\right)\right)\right]_{u=0},
\end{eqnarray}
where the prime denotes the derivative with respect to $u$. As shown in \cite{Hassanain:2011ce}, the reduction of $\lambda$ results in the enhancement of the peak of the spectrum and the shift toward the infrared region. 
Although the DC conductivity and the maximum amplitude of the spectrum are enhanced for smaller $\lambda$ \cite{Hassanain:2012uj}, $\eta_{\gamma}^{SYM}$ increases when $\lambda$ decreases as shown in Fig.\ref{photon_finite_coupling}, where the divergence of $\eta_{\gamma}$ at $\lambda\approx 50$ is due to the breakdown of the perturbative expansion. On the contrary, by using the result in \cite{Muller:2015maa}, where $\eta_{e}^{SYM}/T^3$ is inversely proportional to the ratio of DC conductivity of the SYM plasma to temperature, the shear viscosity of electrons $\eta_{e}^{SYM}$ is suppressed when $\lambda$ decreases as shown in the same plot \footnote{Here only the lepton-quark scattering is considered.}.   
In Fig.\ref{photon_weak_coupling}, we also present the shear viscosity at weak $\lambda$, where the spectral functions and DC conductivity are obtained from the perturbative calculations in the SYM plasma \cite{CaronHuot:2006te} and in the QGP \cite{Arnold:2001ba,Arnold:2001ms,Jiang:2014ura}. Here we cut the results at $\lambda\approx 2.5$, where the $\sigma_c$ of QGP starts to decrease when increasing $\lambda$, which may imply the breakdown the leading-order computation \footnote{The validity of the perturbative calculations for $\lambda\geq 3(g_{YM}\geq 1)$ could be questionable.}. Also, the $\sigma_c$ of the SYM plasma becomes singular at $\lambda=4.5$. It turns out that the increase of $\eta_{\gamma}^{SYM}$ with respect to the decrease of $\lambda$ is as well found in the weakly coupled scenario. Analogously, the decrease of $\eta_{e}^{SYM}$ with respect to the decrease of $\lambda$ is observed. Unlike $\eta^{SYM}_e$ in the SYM plasma, where the low-energy scattering dominates and thus the increase of the DC conductivity of the plasma reduces $\eta^{SYM}_e$, the $\eta^{SYM}_{\gamma}$ is governed by the full spectral function. It is in fact the blue-shift of the spectrum led by the increase of $\lambda$ found in both weakly coupled and strongly coupled scenarios causes the suppression of $\eta_{\gamma}$. Based on the qualitative features of the shear viscosity at weak and strong couplings, we may expect monotonic decrease of $\eta_{\gamma}$ and monotonic increase of $\eta_{e}$ with respect to $\lambda$. 
Note that $\eta^{SYM}_{\gamma}(2.5)\approx 1.4\eta^{SYM}_{\gamma}(\infty)$, which may further suggest that $\eta^{SYM}_{\gamma}$ could almost saturate the value at $\lambda=\infty$ at intermediate $\lambda$. 
  
In the following, we investigate the shear viscosity of photons in the sQGP via holographic QCD. Although the $\mathcal{N}=4$ SYM theory and QCD may have similar features at intermediate temperature, the difference of two theories will become substantial near the critical temperature, where the conformal anomaly of QCD results in the deconfinement transition. In order to break the conformal symmetry in holography, one has to incorporate the running of coupling, which could be characterized by a bulk scalar field with a corresponding potential in the gravity dual in the bottom-up approaches. This type of models may be regarded as an effective theory of QCD in the IR regime. Here we will employ the phenomenological model in \cite{Gubser:2008ny,Gubser:2008yx}, which generates thermodynamic properties similar to the results in lattice QCD. In \cite{Gubser:2008ny,Gubser:2008yx}, the gravitational action in the Einstein frame takes the form
\begin{eqnarray}
S=\frac{1}{2\kappa_5^2}\int d^5x\sqrt{-G}\left[\mathcal{R}-\frac{(\partial\phi)^2}{2}-V(\phi)\right], 
\end{eqnarray}
where $G$ represents the determinant of the spacetime metric $G_{\mu\nu}$ and $V(\phi)$ denotes the scalar-field potential. Here $\kappa_5^2=16\pi^2L/N_c^2$ with $L$ being the AdS radius. The model is usually solved in the Gubser gauge with the following ansatz of the metric,
\begin{eqnarray}
ds^2=e^{2A(\phi)}\left(-h(\phi)dt^2+d{\bf x}^2\right)+e^{2B(\phi)}\frac{d\phi^2}{h(\phi)},
\end{eqnarray}
where the scalar field $\phi$ is set as the fifth coordinate in the bulk. The blackening function $h(\phi)$ should have a simple zero at the horizon $\phi=\phi_h$ and the metric should recover $AdS_5$ geometry near the boundary at $\phi\rightarrow 0$. The temperature and entropy density are given by
\begin{eqnarray}
T=e^{A(\phi_h)-B(\phi_h)}\frac{|h'(\phi_h)|}{4\pi},\quad s=\frac{2\pi}{\kappa_5^2}e^{3A(\phi_h)},
\end{eqnarray}
where the prime above denotes the derivative with respect to $\phi$. In light of \cite{Finazzo:2013efa}, we choose the following potential,
\begin{eqnarray}
V(\phi)=-12\cosh\gamma\phi+b_2\phi^2+b_4\phi^4+b_6\phi^6,
\end{eqnarray}
where $\gamma=0.606$, $b_2=0.703$, $b_4=-0.12$, $b_6=0.0044$, and the AdS radius $L=1$. We will not present the details for solving the metric, which could be found in \cite{Gubser:2008ny}.

\begin{figure}[t]
{\includegraphics[width=7cm,height=5cm,clip]{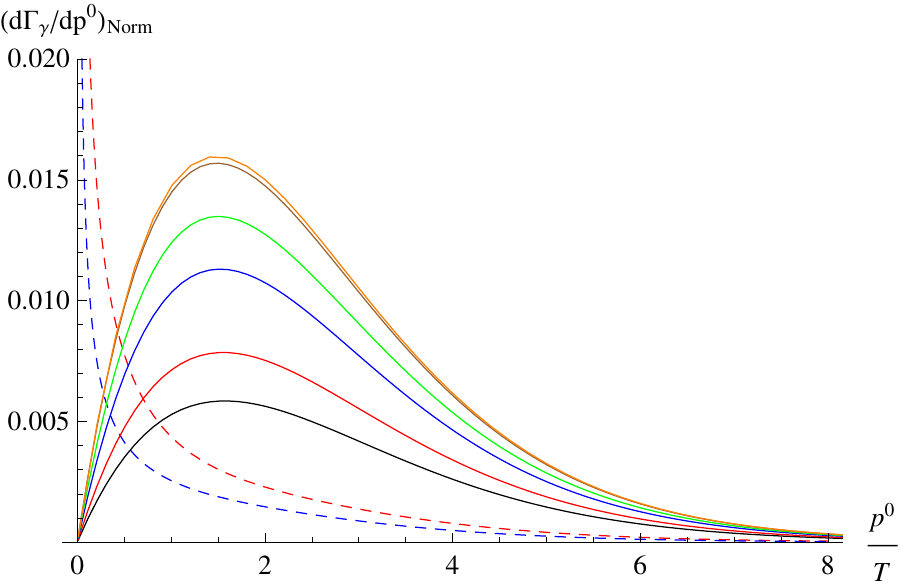}}
\caption{Solid curves from bottom to top correspond to the photoemission rates in holographic QCD in strong coupling at $T=$(1, 1.08, 1.26, 1.54, 3.57, 7.4) $T_c$, respectively. Dashed curves from bottom to top at $p^0/T=2$ correspond to the rates at weak coupling with $\alpha_s=0.05$ and $\alpha_s=0.1$.}\label{photon_spectra}
\end{figure}

\begin{figure}[t]
	{\includegraphics[width=7cm,height=5cm,clip]{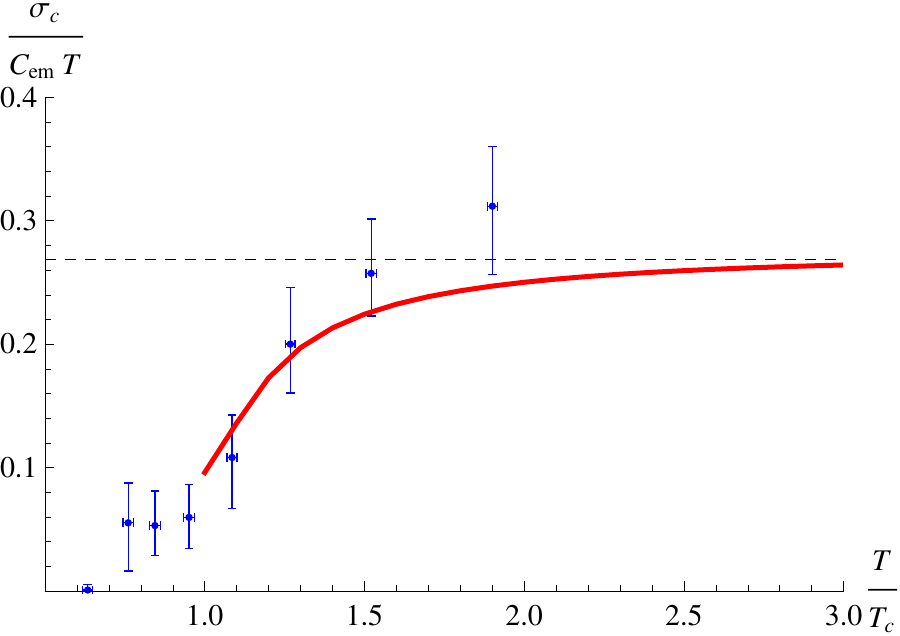}}
	\caption{ The solid red curve corresponds to normalized conductivity obtained from the holographic QCD and the dashed line corresponds to the one from strongly coupled $\mathcal{N}=4$ SYM plasma. The blue points represent the lattice simulation for the light+strange case in \cite{Aarts:2014nba}. Here $C_{em}=2e^2/3$.}\label{DC_cond}
\end{figure}

Now, we have to coupled an U(1) gauge field to the gravitational action. The relevant term which follows the linear response of the U(1) current can be written as
\begin{eqnarray}
S_{F}=-\frac{1}{2\kappa_5^2}\int d^5x\sqrt{-G}\frac{f(\phi)}{4}F^{MN}F_{MN},
\end{eqnarray}
where $F_{MN}$ denotes the field strength. 
We will take
\begin{eqnarray}
f(\phi)=\frac{\text{sech}(a_1\phi)}{g_{5,1}^2},
\end{eqnarray}
which is introduced in \cite{Finazzo:2013efa} to fit the electric susceptibility from lattice simulations for $T<1.5T_c$
\footnote{There are other choices of $f(\phi)$ in \cite{Finazzo:2013efa}, which give rise to similar observables.}
, where $a_1=0.4$ and $g_{5,1}$ is a dimensionless constant associated with the overall amplitude of the observables. More importantly, the choice qualitatively captures the increase of $\sigma_c$ near $T_c$ and the saturation of the SYM result around $T\approx 3.5 T_c$ in comparison with the lattice simulations \cite{Greif:2014oia}.    
We then solve the Maxwell equation numerically with the incoming-wave condition at the horizon. From the standard AdS/CFT prescription \cite{Policastro:2002se}, the trace of the spectral function reads
\begin{eqnarray}
\chi(p^0)=\frac{e^2N_c^2}{16\pi^2}\text{Im}\left[\lim_{\phi\rightarrow 0}\left(f(\phi)\sqrt{-G}G^{\phi\phi}G^{ii}\frac{\partial_{\phi}E_i}{E_i}\right)\right],
\end{eqnarray}
where $E_i=|{\bf p}|A_i$.         
In Fig.\ref{photon_spectra}, we illustrate the thermal-photon spectra at different temperature in strong coupling, where
\begin{eqnarray}
\left(d\Gamma_{\gamma}/dp^0\right)_{\text{norm}}=\frac{d\tilde{\Gamma}^{<}/dp^0}{\alpha_{EM}N_c^2T^3}.
\end{eqnarray} 
Here we extract $T_c$ from the minimum of the speed of sound \cite{Gubser:2008yx} and fix the overall amplitude of the spectrum with a proper choice of $g_{5,1}$ by matching the result at $T=3.54T_c$, where the spectra at higher $T$ start to saturate, and the one from the $\mathcal{N}=4$ SYM plasma. Such a choice yields the conductivity shown in Fig.\ref{DC_cond} in comparison with the one in the SYM plasma \cite{CaronHuot:2006te} at $\lambda=\infty$ and the lattice simulation in \cite{Amato:2013naa,Aarts:2014nba}.  
In Fig.\ref{photon_spectra}, we also show the results from the weakly coupled QCD at high temperature \cite{Arnold:2001ba,Arnold:2001ms}, where we take $N_c=3$ and $N_f=3$. Similar to the scenario in the $\mathcal{N}=4$ SYM plasma, one finds substantial blue-shift of the photon spectrum at strong coupling. In fact, the blue-shift may be foreseen in the long-wavelength limit. Up to the leading order of the small-momentum expansion of the light-like spectral functions, the AdS/CFT calculations yields $(\chi/T^2)\sim p^0/T$, which follows the typical pattern of hydrodynamics, whereas the perturbative QCD and $\mathcal{N}=4$ SYM theory lead to $(\chi/T^2)\sim(p^0/T)^{-1/2}\ln(p^0/T)$ and $(\chi/T^2)\sim(p^0/T)^{-3/2}\ln(p^0/T)$, respectively. 

\begin{figure}[t]
{\includegraphics[width=7cm,height=5cm,clip]{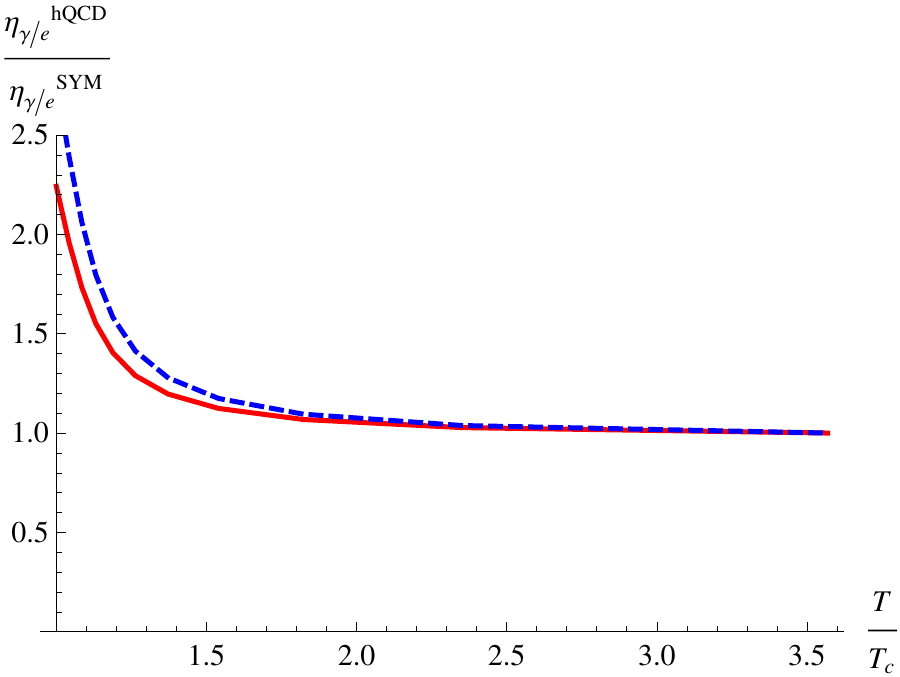}}
\caption{ The solid and dashed curves correspond to the ratio of $\eta_{\gamma}$ and the ratio of $\eta_e$.}\label{eta_ratio_T}
\end{figure}

However, to locate the peak of the spectrum beyond the hydrodynamic region, one has to perform the computations of full spectra.
In the holographic model, we find that the peaks of the spectra at different temperature almost locate at the same momentum scaled by the temperature. Furthermore, the normalized spectra with high temperature far from $T_c$ almost coincide. This finding agrees with the saturation of the ratio of conductivity to temperature found in the same model \cite{Finazzo:2013efa}. After obtaining the photon spectrum, we can employ (\ref{eta_gamma2}) to evaluate the shear viscosity of thermal photons in the holographic QCD $\eta_{\gamma}^{hQCD}$. Since the shape of the spectrum is nearly unchanged by varying $T$, $\eta_{\gamma}^{hQCD}$ scaled by $T^3$ rapidly increases near $T_c$ as shown in Fig.\ref{eta_ratio_T}, which is qualitatively analogous to the behavior of the shear viscosity of thermal leptons $\eta_{e}^{hQCD}$ \cite{Muller:2015maa}.

In summary, we found two salient features of the photon transport in the sQGP. First, albeit it is only shown in the $\mathcal{N}=4$ SYM plasma, thermal photons in the sQGP become more fluid-like when the coupling increases. Second, the shear viscosity of photons in the sQGP increases near the deconfinement transition. 
Nevertheless, both the shear viscosity and relaxation time of thermalized photons depend on the electric conductivity of the QGP. According to \cite{Amato:2013naa}, $\sigma_c$ in the QGP from recent lattice simulations is not very far from the value of the strongly coupled $\mathcal{N}=4$ SYM plasma for $T\sim 2-3 T_c$, which may results in somewhat large $\eta_{\gamma}/s_{\gamma}$ as shown in (\ref{photon_etas}). However, $\sigma_c/T$ could vary drastically based on distinct models and approaches \footnote{See eg.\cite{Greif:2014oia} and the references therein.}. For example, by utilizing $\tau_{\gamma}\approx 1/(2\sigma_c)$, we find $\tau_{\gamma}\approx 5-50$ fm at $T=200$ MeV given that $\sigma_c/T\approx 0.01-0.1$ ranging from lattice simulations to perturbative calculations. 
Although such approximated values of the relaxation time are smaller compared with the time scale of quark epoch in cosmology, it is inconclusive to pin down the exact values of $\eta_{\gamma}/s_{\gamma}$ and $\tau_{\gamma}$ in the cosmic plasma. 
On the other hand, our approach is not subject to the thermalized photons/leptons in the deconfined phase. By knowing the photon-emission rate in the hadron gas, one could further analyze the photon/lepton transport in the nuclear matter. In fact, as shown in the phenomenological study in \cite{Greif:2016skc}, where the DC conductivity of the interacting hadron gas is enhanced at low temperature, which may imply further suppression of lepton shear viscosity in the hadron epoch in cosmology. Nonetheless, the change of photon shear viscosity further requires the full photon spectrum in the hadron gas. The direct or indirect influence
of the photon/lepton shear viscosity in the cosmic plasma on experimental observables should be pursued in the future. 

Furthermore, on the theoretic side, one may incorporate different effects such as a strong magnetic field and pressure anisotropy of the QGP, which may modify the conductivity of the QGP and the production rate of thermal photons \cite{Schenke:2006yp,Patino:2012py,Mamo:2013efa,Wu:2013qja,Jahnke:2013rca,Muller:2013ila} in various conditions. In addition, for simplicity, we neglected chemical potentials and bulk fluctuations in our study. One could further generalize the approach to include nonzero chemical potentials or investigate the bulk viscosity of photons/leptons in non-conformal plasmas. For example, as shown in the holographic model of QCD \cite{Finazzo:2015xwa}, in which the photon-production rate is amplified by baryon chemical potentials particularly near the critical temperature, we may expect the shear viscosity of photons/leptons could be reduced near $T_c$ in the presence of chemical potentials.  
Also, although the imaginary part of the EM-current correlator in the sQGP is sufficient for evaluating the production rate and $\eta_{\gamma}$, the real part also contributes to the self-energy of photons. The modification of the photon dispersion relation in the strongly coupled medium may result in a negative refractive index \cite{Amariti:2010jw,Forcella:2014dwa,Forcella:2014gca}. The connection between the refractive index and shear viscosity of photons at strong coupling could make further impacts on different strongly interacting systems. Moreover, the approach for analyzing the interplay between a perturabtive sector and a non-perturbative sector in this paper can also be applied to the semi-holographic model of QCD recently proposed in \cite{Iancu:2014ava}.

{\em Acknowledgement}: The authors thank A. Mukhopadhyay and P. Romatschke for fruitful discussions. This work was
supported by Grant no. DE-FG02-05ER41367 from the
U. S. Department of Energy and in part by European Union's Seventh Framework Programme under grant agreements (FP7-REGPOT-2012-2013-1) no 316165, the EUGreece
program "Thales" MIS 375734 and was also co-financed by the European Union (European Social Fund, ESF) and Greek national funds through the Operational
Program "Education and Lifelong Learning" of the National Strategic Reference
Framework (NSRF) under "Funding of proposals that have received a positive
evaluation in the 3rd and 4th Call of ERC Grant Schemes" and the RIKEN Foreign Postdoctoral Researcher program.


\end{document}